\documentclass{aa}  
\usepackage{graphicx}
\usepackage{natbib}
\usepackage{txfonts}
\usepackage{amssymb}
\usepackage{verbatim}
\newcommand{\chandra}{{\em Chandra}}
\newcommand{\cdfs}{CDFS}
\begin{document}
%
%   \title{Source counts down to the absolute {\it Chandra} limit}
   \title{The nearest neighbor statistics for X-ray source counts\\
          II. Chandra Deep Field South}

   \author{A. M. So\l tan         \inst{}                }

   \offprints{A. M. So\l tan}

   \institute{Nicolaus Copernicus Astronomical Center,
              Bartycka 18, 00-716 Warsaw, Poland\\
              \email{soltan@camk.edu.pl}                 }

   \date{Received ~~ / Accepted }

  \abstract
   {It is assumed that the unresolved fraction of the X-ray background
    (XRB) consists of a truly diffuse component and a population
    of the weak sources below the present detection threshold.
    Albeit these weak sources are not observed directly, their collective
    nature could be investigated by statistical means.}
   {The goal is to estimate the source counts below the conventional
    detection limit in the \chandra\ Deep Field-South 2Ms exposure.}
   {The source number counts are assessed using the nearest neighbor
    statistics applied to the distribution of photon counts. The method
    is described in the first paper of these series.}
   {The source counts down to $3-4\cdot 10^{-18}$\,cgs in the soft band
    ($0.5-2$\,keV) and down to $2-3\cdot 10^{-17}$\,cgs in the hard
    band ($2-8$\,keV) are evaluated. It appears that in the soft band
    the source counts steepen substantially below $\sim\!10^{-16}$\,cgs. 
    Assuming that the differential slope $b \approx 1.5 - 1.6$ in the range
    $10^{-16} - 10^{-14}$\,cgs, the number of weaker sources indicates
    the slope of $\approx -2.0$. The steepening is not observed in the
    hard band.}
   {Steepening of counts in the soft band indicates a new population of
    sources. A class of normal galaxies at moderate redshifts is
    a natural candidate.}


   \maketitle
%
%________________________________________________________________

\section{Introduction}

Source number counts are investigated in one of the deep \chandra\
pointing, viz. the \chandra\ Deep Field South (CDFS). An original
method based on the nearest neighbor statistics (NNST) has been
used. In the first paper \citep[hereafter~SI]{soltan10}
we presented the method and demonstrated its efficiency for
the source number counts analysis. It was applied to the
\chandra\ exposure of $\sim\!460$\,ks in the Groth Strip.
The  NNST allowed us to determine the $N(S)$
relationship for sources generating merely $\sim\!2$ counts,
i.e. roughly an order of magnitude below a standard threshold for
a detection of discrete point-like sources. Application of the
NNST to the \cdfs\ shifts the sensitivity limit below
$4\cdot 10^{-18}$\,cgs, in the $0.5-2$\,keV band, still a factor
of $\sim\!3$ below the present deepest number counts determinations
\citep[hereafter~GNL]{georgakakis08}.

A comprehensive discussion of the present method is given in SI.
Here only a general framework is sketched. From the point of view
of spatial characteristics, the counts collected in the focal plane
of the X-ray telescope are arranged into two populations. The
first, randomly distributed, includes the X-ray photons generated
by a truly diffuse XRB and locally scattered X-rays of various
origin as well as the particle background. The weak sources which
generate in the exposure exactly one count each also contribute to
this category, if they are distributed randomly within the field of
view.  The second class is generated by the discrete sources
producing at least two photons. Counts of this class are
distributed in clumps defined by the point spread function (PSF) of
the X-ray telescope.

One should expect that a distribution of the nearest neighbors for
both classes of counts is different. On the average, the counts
produced by sources have closer neighbors than the non clustered
counts.

The observed distribution of the nearest neighbors results from the
relative contribution of the randomly distributed counts, $n_1$,
and the number of sources, $N_k$, producing $k$ photons each, where
$k = 2, 3, ..., k_{\rm max}$, with $k_{\rm max}$ representing the
strongest source in the field. The total number of counts is a sum
of both constituents:

\begin{equation}
n_t = n_1 + \sum_{k=2}^{k_{\rm max}} k\cdot N_k\,.
\end{equation}

We define $P(r)$ as the probability that the distance to the nearest
neighbor from the randomly picked count is greater then $r$.
The relationship between the source population $N_k$ and the
probability $P(r)$ is given in SI:

\begin{equation}
\label{s4}
\frac{n_1}{n_t}\,P(r|1)\: +\: \sum_{k=2}^{k_{\rm max}}\;
    \frac{n_k}{n_t}\;P(r|1)\:{\cal P}(r|k)\: =\: P(r)\,,
\end{equation}
where $P(r|1)$ is the probability that the distance to the nearest
neighbor from the randomly selected {\it point} (not count) exceeds
$r$, $n_k = k\cdot N_k$ is the total number of photons in the field produced
by sources generating $k$ photons each, and ${\cal P}(r|k)$ describes
the nearest neighbor probability within a cluster of $k$ counts
generated by a single source. The probability ${\cal P}(r|k)$ is fully
defined by the PSF.

The numbers of sources $N_k$ depend on the source number counts
$N(S)$. The relationship between the source flux, $S$, and the
number of actually detected source counts, $k$, is given by the
Poissonian distribution:
\begin{equation}
p(k|S) = \frac{e^{-s}\,s^k}{k!}\,,
\end{equation}
where $p(k|S)$ is the probability that source with flux $S$ generates
$k$ counts, while $s$ is the expected number of counts, or it
is the source flux expressed in the units of counts. The
flux $s$ in the ACIS\footnote{See http://asc.harvard.edu/ciao.}
counts is related to the flux in physical units, $S$, by:

\begin{equation}
\label{def_cf}
s = S / {\rm cf}\,,
\end{equation}
where ${\rm cf}$ is the conversion factor which has units of
``erg\,cm$^{-2}$\,s$^{-1}/{\rm count}$'' and is related to the parameter
``exposure map'' defined in a standard processing of the ACIS data.
For the real observations, both the cf and exposure map are functions
of the position. The present analysis is restricted to the area
where the cf variations are small (see below). A question of 
the cf variations over the field of view is discussed in SI.

For the power law number counts $N(s) = N_o\,s^{-b}$, Eq.~\ref{s4}
takes the form:
\begin{equation}
\label{final}
\frac{N_o}{n_{\rm t}}\, P(r|1)
     \sum_{k=2}^{k_{\rm max}} \frac{\Gamma(k-b+1)}{\Gamma(k)}\:
     \left[1-{\cal P}(r\!\mid\!k)\right] = P(r\!\mid\!1) - P(r)\,.
\end{equation}
For different parametrization of the source counts the
Eq.~\ref{final} could be modified, in particular for the broken power
law:
\begin{equation}
%\[
\label{bpl}
N(s) = N_o \left(\frac{s}{s_o}\right)^b\,,\hspace{10mm} b = \left\{ 
                 \begin{array}{ccc}
                  b_1 & \mbox{for} & s \ge s_o \\
                  b_2 &            & s <   s_o
                 \end{array}
 \right.
%\]
\end{equation}
the $\Gamma$ function in the numerator is replaced by the appropriate
combination of the incomplete $\Gamma$ functions. 

Since we are interested in the distribution of weak sources, all the
discrete sources strong enough to be isolated using the standard
methods should be extracted from the observations. After the removal
of bright sources, the $k_{\rm max}$ value corresponds to the weakest
sources which could be unmistakably recognized as individual objects.
In derivation of Eq.~\ref{final} it is assumed also that the exposure
is sufficiently deep to use the functional form of $N(s)$ rather than
the actual number of sources $N_k$.

A standard method to estimate the number counts of weak sources is
based on the fluctuation analysis. The sources in the observation
area increase fluctuations of the count number in the detection cell
above that expected for the Poissonian distribution. The fluctuation
enhancement is dominated by the strongest sources, while it is very
weakly affected by the faint sources which contribute only a few
counts. This is because the fluctuations amplitude is related to
the second moment of the count distribution. 

The NNST weights the contribution of
weak and strong sources more evenly: a deviation from the random
distribution defined as the right-hand side of Eq.~\ref{final}
depends linearly on the number of counts $n_k$. Thus, the NNST
appears to be a relatively sensitive tool to quantify the
contribution of counts produced by weak sources and can be
efficiently applied to assess the $N(S)$ relationship at the low flux
end.

The organization of the paper is as follows. In the next section the
observational material is described and the computational details
including questions related to the PSF are given.  Results of the
calculations, i.e.  estimates of the source counts below the nominal
sensitivity limit in the CDFS are presented in Sect.~3. The results
are summarized and discussed in Sect.~4.

\section{Observational material \label{observations}}

In the present paper we analyze the counts collected in the
\chandra\ ACIS-I chips 0\,-\,3.  The CDFS was observed with the ACIS
detector in several "sessions". Although the observations span a
period of more than $7$ years, the data have been processed in a
uniform way with the recent pipeline processing versions. The details
of $20$ observations used in the present paper are given in
Table~\ref{obs_log}. All the exposures have been scrutinized with
respect to the background flares and only the ``good time
intervals'' were used in the subsequent analysis. The data
have been split into two energy bands: S -- soft ($0.5-2$\,keV)
and H -- hard ($2-8$\,keV).

\begin{table}
\caption{Log of the CDFS observations used in the paper}
\label{obs_log}
\centering
\begin{tabular}{rcccr}
\hline\hline
\noalign{\smallskip}
Obs.&\multicolumn{2}{c}{
             Observation and processing}&Processing & Exposure \\
 ID &\multicolumn{2}{c}{dates}          &version    & time [s] \\
\hline
\noalign{\smallskip}
 441&  2000-05-27  & 2007-05-23 &  7.6.10~~~ &   56600 \\
 582&  2000-06-03  & 2007-05-23 &  7.6.10~~~ &  132150 \\
2405&  2000-12-11  & 2007-06-20 &  7.6.10~~~ &   57100 \\
2312&  2000-12-13  & 2007-06-20 &  7.6.10~~~ &  125100 \\
1672&  2000-12-16  & 2007-06-20 &  7.6.10~~~ &   96150 \\
2409&  2000-12-19  & 2007-06-21 &  7.6.10~~~ &   69850 \\
2313&  2000-12-21  & 2007-06-22 &  7.6.10~~~ &  131900 \\
2239&  2000-12-23  & 2007-06-22 &  7.6.10~~~ &  132600 \\
8591&  2007-09-20  & 2007-09-21 &  7.6.11.1  &   45800 \\
9593&  2007-09-22  & 2007-09-27 &  7.6.11.1  &   46100 \\
9718&  2007-10-03  & 2007-10-06 &  7.6.11.1  &   49850 \\
8593&  2007-10-06  & 2007-10-08 &  7.6.11.1  &   49250 \\
8597&  2007-10-17  & 2007-10-24 &  7.6.11.2  &   59650 \\
8595&  2007-10-19  & 2007-11-07 &  7.6.11.2  &  116850 \\
8592&  2007-10-22  & 2007-11-01 &  7.6.11.2  &   87750 \\
8596&  2007-10-24  & 2007-11-14 &  7.6.11.2  &  116600 \\
9575&  2007-10-27  & 2007-11-14 &  7.6.11.2  &  110150 \\
9578&  2007-10-30  & 2007-11-30 &  7.6.11.2  &   39000 \\
8594&  2007-11-01  & 2007-11-14 &  7.6.11.2  &  143300 \\
9596&  2007-11-04  & 2007-11-20 &  7.6.11.2  &  116600 \\
    &   &  &\multicolumn{2}{r}{Total exposure~~1782350}\\
\hline
\end{tabular}
\end{table}

\begin{table}
\caption{Energy bands and conversion factors}
\label{exposures}
\centering
\begin{tabular}{lcllll}
\hline\hline
\noalign{\smallskip}
\multicolumn{2}{c}{Energy band}&
                   \multicolumn{4}{c}{Conversion factors$^\dagger$} \\
\multicolumn{2}{c}{~~~~~~~~~[keV]}
               &  Average  & ~~rms     &  minimum    &   maximum  \\
\hline
\noalign{\smallskip}
S  &$0.5 - 2$  & ~~$3.701$ &  $0.205$  & ~~~$3.232$ & ~~~$4.218$ \\
H  & $2   - 8$ & ~~$14.30$ &  $0.85~$  & ~~~$12.69$ & ~~~$16.28$ \\ 
\hline
\noalign{\smallskip}
\multicolumn{6}{p{85mm}}{$^\dagger$ The conversion factor (cf) has units of
$10^{-18}\,{\rm erg\,cm}^{-2}\,{\rm s}^{-1} / {\rm count}$.}
\end{tabular}
\end{table}

\subsection{The exposure map \label{expmap}}

The observations, listed in Table~\ref{obs_log} were merged to create
a single count distribution and exposure map. A circular area covered
by all the pointings with a relatively uniform exposure, centered at
${\rm RA} = 3^{\rm h}32^{\rm m}$, ${\rm Dec} = -27\deg 49\arcmin$ with
radius of $5\farcm0$ in the S band and $4\farcm0$ in the H band has
been selected for further processing. The exposure map of the
individual observation resulting from various instrumental
characteristics has a complex structure. The exposure map of the
merged observation exhibits significant variations too, although it is
more uniform than the individual components.  To reduce further the
variations of the exposure map over the investigated area, a threshold
of the minimum exposure has been set separately for both energy band.
Pixels below this threshold have not been used in the calculations.

A threshold has been defined at $77$\,\% of the maximum value of the
exposure map in the S band and $78$\,\% in the H band. As a result, in
both energy bands the maximum deviations of the exposure from the
average value do not exceed $15$\,\% and the exposure rms over the
investigated areas fall below $6$\,\%. In Table~\ref{exposures} the
conversion factors corresponding to the relevant amplitudes of the
exposure maps are given. In the calculations ``from counts to flux'' a
power spectrum with a photon index $\Gamma_{\rm ph} = 1.4$ was assumed
\citep{kim07}.

Variations of the conversion factor, ${\rm cf}$, over the investigated
area alter the source fluxes via Eq.~\ref{def_cf} and, consequently,
the source counts $N(S)$. However, it is shown in SI that in the
linear approximation variations of the exposure maps and the
conversion factor do not affect the probability distributions
$P(r\!\mid\!1)$ and  $P(r)$. Thus, the restrictive limits imposed on
the cf fluctuation amplitude ensure that the NNST should yield reliable
results.

\subsection{The count selection}

A single cosmic ray can induce in the ACIS CCD detector a series of
$2$ or more ``events''. This well recognized feature\footnote{See
http://cxc.harvard.edu/ciao/why/afterglow.html for details.}, known as
``afterglow'', generates spurious weak sources in the data and
potentially could affect the present investigation. Fortunately, as
indicated in SI, the afterglow counts span short time intervals as
compared to the exposure times of all the observations.  The material
has been scrutinized with respect to the afterglows and events
identified with that phenomenon have been removed from the observation.

Strong sources were localized in the field using the
\citet{giacconi02} catalog based on 1Ms exposure. Around each
cataloged source position a radius $r_{85}$ encircling $85$\,\% of the
point source counts has been calculated using the local PSF parameters
(see below).  Then, number of counts within $r_{85}$ was obtained, and
- by subtracting the background counts - the net counts $k_{85}$ were
assessed for the each source. Since the relative variation of the exposure
within the field of view (fov) are small, the average background was
assumed for the entire field.
The threshold counts $k_{\rm max}$ characterizing the
completeness limit of the catalog is not well defined, and a range of
$k_{\rm max}$ between $20$ and $45$ were applied in the
analysis in both energy bands. For given $k_{\rm max}$, the source
has been excluded from further processing  if $k_{85} > 0.85\cdot
k_{\rm max}$. To assure the removal of the source counts in the PSF
wings, the rejection area was a circle with radius
$r_{\rm rem} = 4\cdot r_{85} + 4\arcsec$.

In the standard processing of the ACIS data the discrete pixel
coordinates are randomized over the square pixel size of $0\farcs492$
a side. Accordingly, the count separations are subject to
randomization at the pixel size scale. To assess the effect of
randomization on the NNST probability distributions, a set of $12$
``observational`` data was generated by randomization of count
position within pixels using the original event files with
non-randomized (integer) count coordinates.

\subsection{The Point Spread Function \label{psf_sec}}

The ${\cal P}(r|k)$ probability has been calculated by means of the
Monte Carlo method using the model PSF. The procedure to construct
the PSF suitable to the present investigation is described in
details in SI.

\begin{figure}
\resizebox{\hsize}{!}{\includegraphics{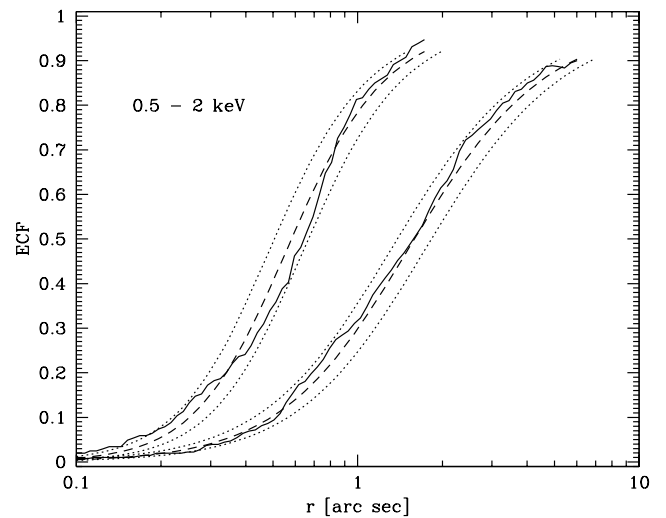}}
\caption{Encircled count fraction (ECF) in the S band as a function
of a distance from the count centroid. Example distributions are shown
for $2$ sources at $2\farcm 1$ and $5\farcm 9$ from the field center.
Solid curves -- the observed count distributions, dashed curves -- fits
obtained using Eqs.~\ref{psf} and \ref{psf_par}, dotted curves --
the ECF distributions differing from the best fit by $\pm 15$\,\%.}
\label{ecf_fit_1}
\end{figure}

\begin{figure}
\resizebox{\hsize}{!}{\includegraphics{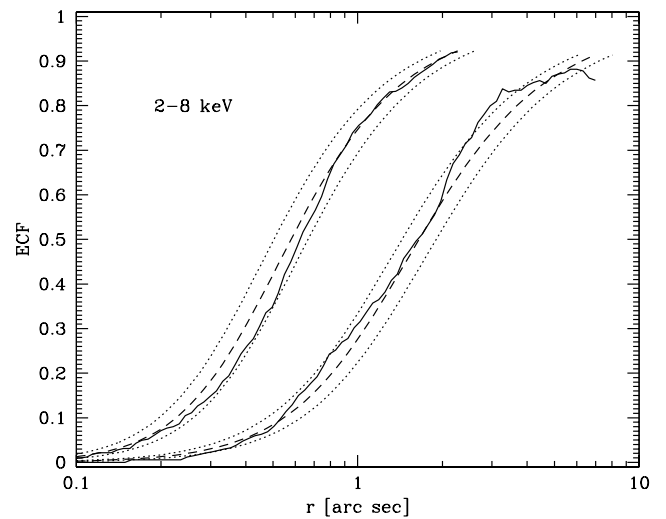}}
\caption{Same as Fig.~\ref{ecf_fit_1} for the H band.}
\label{ecf_fit_4}
\end{figure}

The \chandra\ X-ray telescope PSF is a complex function of source
position and energy \citep[e.g.][]{allen04}. To compute the ${\cal
P}(r|k)$ probability which describes the nearest neighbor distribution
in the entire data set, we need to generate the PSF appropriately
averaged over the field of view. A tractable method to obtain the
${\cal P}(r|k)$ was to find an analytic approximation for the
encircled count fraction (ECF) for a point source as a function of the
distance from the field center. To reproduce the count distribution
for a single source, we have used the function of the form
\begin{equation}
\label{psf}
ECF(<r) = \frac{r^\alpha}{z + r^\alpha + y\cdot r^{\alpha/2}}
\end{equation}
with free  parameters $\alpha$, $z$, and $y$. A shape of the PSF
depends strongly on the distance from the optical axis of the
telescope. For a single observation the optical axis is shifted from
the geometrical center of the ACIS chips 0\,-\,3. However, in the
merged data of $20$ pointings the axis appropriate for the PSF
modeling is not well defined and it was assumed that the variations
of the PSF shape are symmetrical with respect to the fov center.
By fitting the $\alpha$, $z$, and $y$ to the ECF
distributions of a number of sources scattered over the
entire fov it was found that variations of these parameters
can be conveniently parametrized by the distance from the field center
$\theta$. It was assumed that:

\begin{equation}
\label{psf_par}
\alpha = a_\alpha \cdot \theta + b_\alpha \\
y = a_y \cdot \theta + b_y \\
\log z = a_z \cdot \theta + b_z\,,
\end{equation}
where $a_s$ and $b_s$ ($s = \alpha$, $y$, $z$) are six parameters
which are substituted in Eq.~\ref{psf} and simultaneously fitted to
the observed distribution of counts in the several dozen strongest
sources.

In Figs.~\ref{ecf_fit_1} and \ref{ecf_fit_4} examples of the resultant
fits to the observed distribution are shown in both energy bands.
Although the fitting procedure provides sensible and functional
representation of the PSF over the fov, it is difficult to assess the
impact of the potential systematic errors generated by the present
approximation on our final results.  To control the systematics, we
have constructed two model ${\cal P}(r|k)$ distributions using the ECF
functions systematically wider and narrower by $15$\,\% as compared to
the best fit.

Examples of the ECFs differing from the best fit by $15$\,\% are
shown in Figs.~\ref{ecf_fit_1} and \ref{ecf_fit_4} with the dotted
curves.  Albeit deviations of individual fits are quite large, the
$\pm 15$\,\% ECF envelopes undoubtedly encompass the systematic errors
produced by Eqs.~\ref{psf} and \ref{psf_par}.

In the Monte Carlo computations of ${\cal P}(r\!\!\mid\!\!k)$ a
population of $10^8$ ``sources'' of $k = 2, 3, ..., k_{\rm max}$
counts were distributed randomly over the investigated area. The
distribution of counts within each source was randomized according to
the model ECF. Then, for each source a distribution of the nearest
neighbor separations was determined and used to obtain the
corresponding amplitudes of ${\cal P}(r\!\!\mid\!\!k)$. The procedure
has been executed for the best fit and $\pm 15$\,\% ECF distributions.

\begin{figure}
\resizebox{\hsize}{!}{\includegraphics{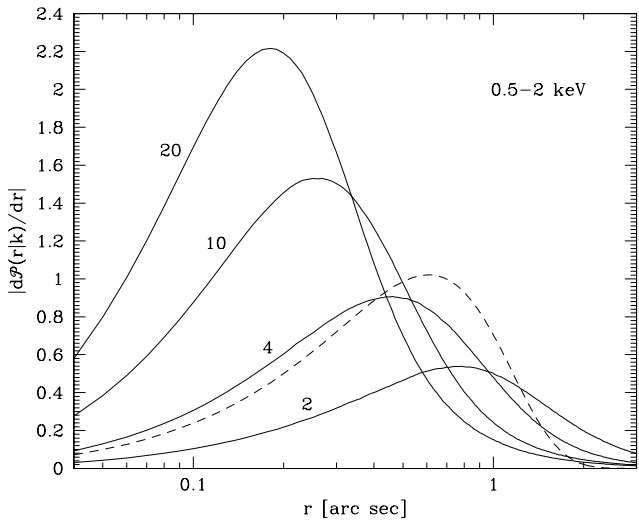}}
\caption{The probability density of the distance to the nearest
neighbor in the S band for counts produced by sources generating
$k = 2$, $4$, $10$, and $20$ counts. The dashed curve shows the
expected distribution for the pure random count distribution.}
\label{prk}
\end{figure}

In Fig.~\ref{prk} the probability densities based on the integral
distributions ${\cal P}(r\!\!\mid\!\!k)$ in the S band are shown
for several values of $k$. To visualize more
clearly details of the relevant distributions, probability densities,
i.e. $|d{\cal P}(r|k)/dr|$ rather than ${\cal P}(r\!\!\mid\!\!k)$
are plotted. The dashed curve shows the probability density of the
nearest neighbors distances for the random distribution.

Conspicuously, due to a high average count density, the nearest
neighbor of the photon generated by the sources producing $k = 2$
counts, is less likely to originate from the same source rather than to
be a chance coincidence with the unrelated event. It shows the natural
limitations of the method. The NNST can be efficiently used for the
investigation of the weak sources if they are sufficiently
numerous to significantly modify the number of the nearest neighbors
observed for the random distribution.

\section{The source counts}

\subsection{The soft band} 

Using the selection criteria given in  Sect.~\ref{observations},
the accepted area and the total number of counts in the soft band
amount to to $71.8$ sq. arcmin and $158\,795$, respectively. After
the removal of strong sources according to the procedure described
above, the area is reduced to $66.5$ sq. arcmin and the number of
counts to $110\,996$ for $k_{\rm max} = 45$ and to $65.4$ sq.
arcmin and $108\,778$ counts for $k_{\rm max} = 30$.  The average
density of counts amounts to $\sim 0.463$ per sq. arcsec.  and the
average distance to the nearest neighbor for the random
distribution is equal to $0\farcs736$. 

All the calculations have been performed in a similar way as in SI.
The nearest neighbor distributions were calculated separately for $12$
data sets obtained by randomization of events within pixels.
Analogously, the distribution of distances between the random points
and the data were obtained. Then, these distributions were used to
calculate the $P(r)$ and $P(r|1)$ probabilities. Taking advantage of a
wide range of separations $r$ over which the probability distributions
were determined, Eq.~\ref{final} was rewritten using the
differential probability distributions $\Delta P(r) = P(r) -
P(r+\Delta R)$ (and $\Delta P(r|1)$ alike). Accordingly, the
Eq.~\ref{final} has been replaced by a set of equations for the 
consecutive values of $r$ and the best estimate of the slope $b$
was found by minimizing the $\chi^2$ of the fit. In our calculations
we used the separations range $0 < r < 2\arcsec$ and $\Delta r = 0\farcs1$.

\begin{figure}
\resizebox{\hsize}{!}{\includegraphics{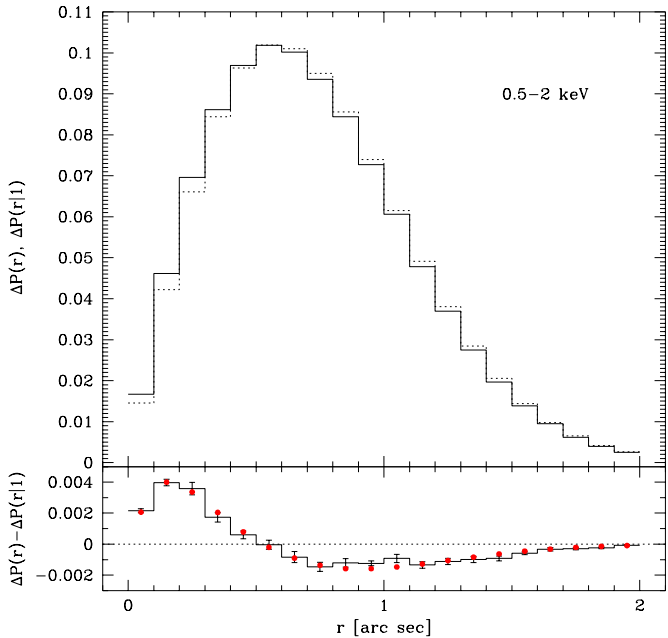}}
\caption{Upper panel: the nearest neighbor probability distributions
binned with
$\Delta r = 0\farcs1$ for the observed counts (solid histogram),
and between random points and the observed counts (dotted histogram).
Lower panel: the difference between both distributions;
the error bars show $1\sigma$ uncertainties; dots - the best fit
(see text for details).}
\label{nnpr}
\end{figure}

Previous investigations of the deep \chandra\ fields, e.g.
\citet{kim07} and GNL, provide essentially consistent
assessments of the $N(S)$ counts above the detection threshold for the
discrete sources.  In particular, in the interesting flux range
GNL approximate the number counts by a power law
with the slope of $-1.58$.  To conform the present investigation to
the observed counts at the bright end, we assume that the $N(s)$
relationship defined in Eq.~\ref{bpl} above $s_o = 20$ counts matches
exactly the GNL model.  Thus, the only parameter to
be determined using the set of equations generated by Eq.~\ref{bpl} is
the slope $b_2$ at the low flux end.

In the upper panel of Fig.~\ref{nnpr} the probability distributions
$\Delta P(r)$ and $\Delta P(r|1)$ are shown for $k_{\rm max} = 40$.
The $\Delta P(r)$ histogram is the average of of $12$ realizations of
the pixel randomization routine. The difference of both distributions
is shown in the lower panel. The error bars represent the rms scatter
between $12$ randomized observations.  The dots show the average of
$12$ best fit solutions obtained using the NNST. The analogous
distributions constructed for several values of $k_{\rm max}$ between
$20$ and $45$ provided qualitatively similar results.

Flux $s_o = 20$ counts marking the slope change
corresponds to $S = 7.84\cdot 10^{-17}$\,cgs. The best fit
slope below the power law break $b_2 = -2.02^{+0.11}_{-0.08}$,
where the errors represent $1\sigma$ statistical uncertainties
(for the full discussion of uncertainties see below).
This result is compared in Fig.~\ref{counts_b1} with the source
counts presented by GNL and our recent estimate
in SI based on the shallower exposure in the AEGIS field.

\begin{figure}
\resizebox{\hsize}{!}{\includegraphics{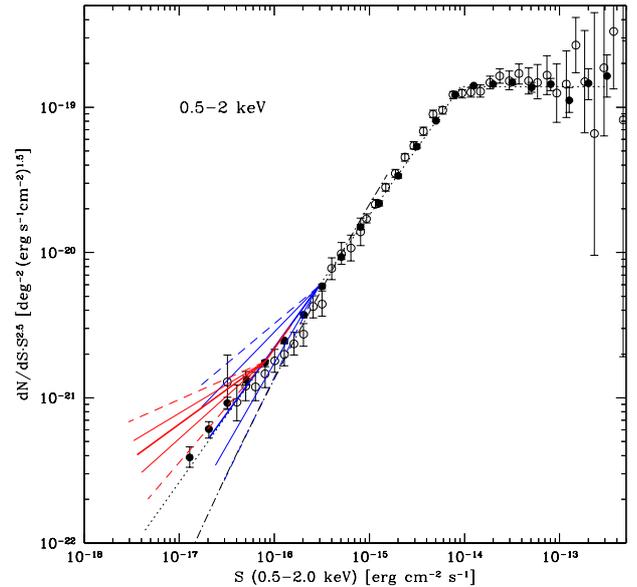}}
\caption{Differential number counts  in the $0.5-2$\,keV band
normalized to the Euclidean slope.
The data points and the dotted line are taken from
\citet{georgakakis08}; dot-dash curve shows the AGN model by
\citet{ueda03}. The right bundle of lines exiting from the
point at $S=3\cdot 10^{-16}$\,cgs represents the solution
obtained in SI based on the pointing at the AEGIS field;
the left bundle shows the present results based on CDFS:
thick line - the best fit solution, thin lines -
$1\sigma$ statistical uncertainty, dashed lines - maximum
total uncertainty range including the potential systematic
errors generated by the approximations in the PSF fitting.}
\label{counts_b1}
\end{figure}

The data points in Fig.~\ref{counts_b1} are based on a large number of
\chandra\ pointings, including CDFS.  The full dots in
Fig.~\ref{counts_b1} show the GNL measurements and
the open circles -- the \citet{kim07} results.  Over a wide flux range
the GNL counts in fact follow the power law with the
slope $-1.58$. Below the flux $S \approx 6\cdot 10^{-17}$\,cgs
GNL notice that the number counts seem to be steeper,
though the deviation from the power law is rather modest.  In the
range of $6\cdot 10^{-17} - 3\cdot 10^{-16}$\,cgs, the number counts
determined by \citet{kim07} run slightly below the
GNL data.

Solid and dashed lines covering fluxes between
$\sim 2\cdot 10^{-17}$ and $3\cdot 10^{-16}$\,cgs are taken
from SI and show the best fit and the statistical uncertainty
as well as the total uncertainties. The present investigation
covers the flux range
of $\sim 3\cdot 10^{-18} - 1.5\cdot 10^{-16}$\,cgs
and in Fig.~\ref{counts_b1} is represented by the best
solution line and lines defining the uncertainty ranges (see
below).

The SI solution in a good agreement with both the
GNL and \citet{kim07} data. Although, the SI
estimate are consistent with the results available in the
literature, the relevant slope uncertainties are
uncomfortably high and do not constrain strongly the $N(S)$
relationship. The present results also are not highly
restrictive. Nevertheless, the acceptable slopes seem to be
distinctly steeper than those by GNL.

\subsection{Error estimates}

\begin{figure}
\resizebox{\hsize}{!}{\includegraphics{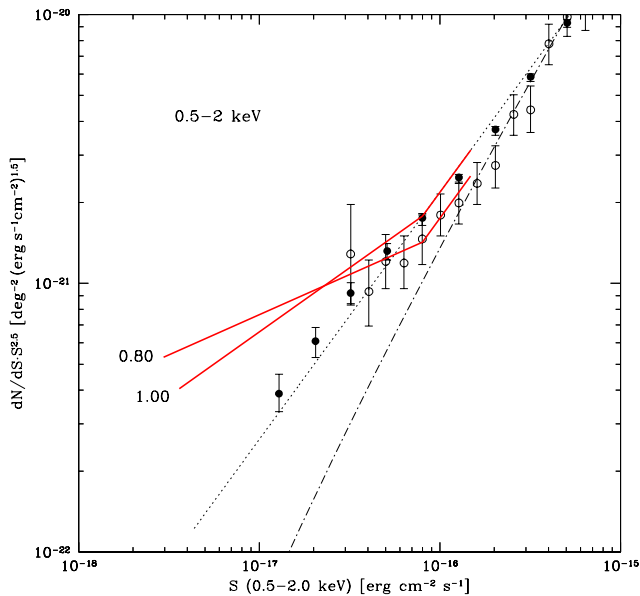}}
\caption{Low flux end of the differential number counts in the
$0.5-2$\,keV band normalized to the Euclidean slope.
The data points, the dotted and dot-dash curves as in
Fig.~\ref{counts_b1}. Broken solid lines: the NNST solutions
for the $N(S)$ with the bright end normalization according to
\citet{georgakakis08} - label "$1.00$", and the normalization
reduced by $20$\,\% - label "$0.80$".}
\label{counts_b1_n0}
\end{figure}

The best estimates of slope in SI and the present results are shown
with the thick solid lines. The uncertainties introduced by the
statistical character of the nearest neighbor method are indicated by
the thin lines. These uncertainties result from variations of the
nearest neighbor probability distributions produced by the
randomization of counts within pixels \footnote{The minimum and
maximum values of $b$ in $12$ data sets are $1.78$ and $2.15$.}.

The systematic errors affecting the investigation are probably
dominated by the inaccuracies in the calculations of the ${\cal
P}(r|k)$ probabilities. These uncertainties have been accounted
for using the ``extreme'' ECF functions described in the
Sect.~\ref{psf_sec}. A set of $12$ solutions has been obtained
using the ${\cal P}(r|k)$ distribution derived from each of the
side ECF. Then, the average values of the slope and the
respective rms amplitudes in both sets were calculated.  The
dashed lines in Fig.~\ref{counts_b1} show the uncertainty range
implied by these calculations, assuming the joint effect of the
systematic errors and the rms scatter.  This estimate of the
``total'' error is highly conservative. It is obtained by simple
addition of statistical uncertainty and the systematic errors
assuming their highest ``reasonable'' values.

A question of the exposure variations over the fov is discussed in SI.
For the power law counts, variations of the exposure map
generate errors in the $N(S)$ which one can express as variations
of the count normalization $N_o$.  
It is shown that constraints imposed in our investigation on the
amplitude of the exposure map variations strongly restrict the
magnitude of the equivalent fluctuations of $N_o$. In effect, the
small exposure map variations do not introduce substantial systematic
uncertainties of the slope determination.

Equation~\ref{final} explicitly involves both parameters which
define the source counts: the slope $b$ and the normalization
$N_o$.  In our calculations only the slope was estimated while the
normalization was fixed. Still, one can obtain a formal solution
for those parameters. Unfortunately, the best estimates of $b$ and
$N_o$ found by a simultaneous fitting are highly correlated. This
is because the NNST is affected predominantly by the total number
of close pairs. Thus, a quality of the fit depends on the proper
combination of $b$ and $N_o$, rather than on each parameter
separately. In Fig.~\ref{counts_b1_n0} two best fit solutions are
shown for two different bright end normalization $N_o$. The broken
line labeled "$1.00$" is the same as in Fig.~\ref{counts_b1}, while
the line labeled "$0.80$" shows the counts with the $N_o$ reduced
by $20$\,\%. 

\subsection{The hard band}

The NNST is basically used to calculate the excess of the close photon
pairs as compared to the number of pairs expected for the random
distribution. High overall count density in the $2-8$\,keV band
significantly limits the efficiency of the NNST method for the
weak source investigation. After the removal of strong sources,
the average distance to the nearest neighbor for the random
distribution amounts to just $0\farcs51$ and the NNST applied
to the $5\arcmin$ radius fov has not
produced any meaningful constraints on the $N(S)$ slope.

\begin{figure}
\resizebox{\hsize}{!}{\includegraphics{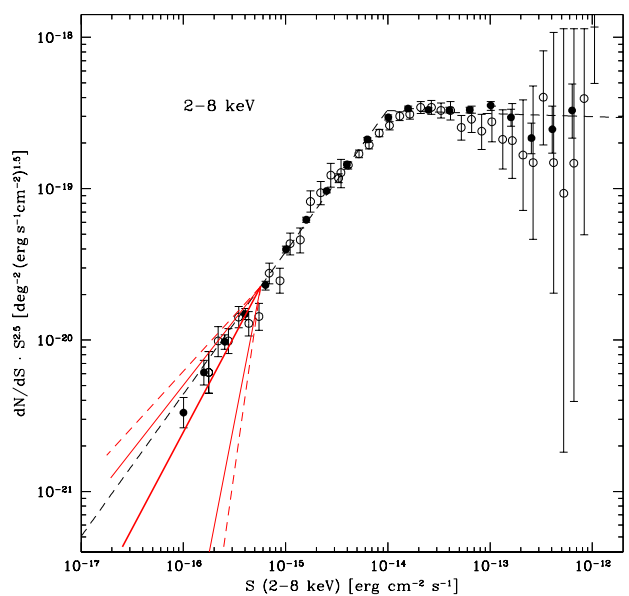}}
\caption{Differential number counts in the $2-8$\,keV band
normalized to the Euclidean slope. The data points and the dashed
line are constructed using the $2-10$\,keV band
from \citet{georgakakis08}; the thick solid line - the
best fit solution, thin lines - $1\sigma$ statistical uncertainty
range, dashed lines - maximum
total uncertainty range including the potential systematic
errors generated in the PSF fitting.}
\label{counts_b4}
\end{figure}

In order to improve the S/N ratio we confined our calculations to the
central area of $4\arcmin$ radius where the PSF is relatively narrow.
The total number of counts within this limited field amounts to
$179371$.  After the removal of sources generating more than $k_{\rm
max} = 40$ the number of counts is reduced to $144632$. Our slope
estimate and its statistical uncertainty $b = -1.23^{+2.28}_{-0.53}$
are consistent with the available source counts. Unfortunately,
the constraints
imposed on $N(S)$ still are not restrictive, particularly the lower
slope limit is insignificant. Nevertheless, the NNST rather strongly
excludes any substantial steepening of the counts, contrary to the
result obtained for the S band. This is shown in Fig.~\ref{counts_b4},
where the present results are confronted with the available
observational material. The points with the error bars are drawn using
the $2-10$\,keV data presented by GNL. The power
spectrum with a photon index of $-1.4$ was assumed to convert fluxes
from the $2-10$\,keV band to our H band.  The NNST solution and
$1\sigma$ statistical uncertainties are shown with the solid thick line
and two thin lines, respectively.  The dashed lines indicate the
combined effect of the maximum potential systematic errors and the
statistical noise.  The systematic errors have been assessed in a
similar way as in the S band. The width of the best fit PSF was altered
by $\pm 15$\,\% to accommodate for conceivable systematic deviations
of the analytic fits from the actual PSF (see Fig.~\ref{ecf_fit_4}).
Then, the new ${\cal P}(r|k)$ probabilities, based on the modified
PSFs, were derived and used in the subsequent calculations.

\subsection{Discussion}

Our slope estimate below $S \approx 7\cdot 10^{-17}$\,cgs is
substantially steeper than the recent estimates by
GNL.  Although these authors also observe the
steepening of counts at the lowest attainable flux levels, their
slope change is distinctly smaller and the discrepancy between our
results remains unexplained. Both the NNST and the GNL
approaches make full use of the Poissonian character of the count
distribution produced by the individual source. Nevertheless, both
methods are distinctly different. GNL assess the source presence
by counting the events within the detection cell, while
here we analyzed the NN distances between the events.
The NNST method has been tested in SI,
but it should be considered still as a new tool, and  one cannot
exclude that unrecognized systematic errors have influenced the
present result. Hopefully, the recent \chandra\ 4Ms observation of
the CDFS would help to clarify this question.

The integral source counts are constrained by the amplitude of the
extragalactic XRB component. A varying galactic contribution to the
total signal makes the estimates of the extragalactic part in the S
band somewhat uncertain. As a reference figure we adopt the XRB
assessment by \citet{moretti03} of $f_S = (7.53 \pm 0.35)\cdot
10^{-12}$\,erg\,s$^{-1}$cm$^{-2}$deg$^{-2}$.  The counts described
by the GNL model integrated above $S = 7.5\cdot 10^{-17}$\,cgs
($\equiv 20$ counts in the present investigation)
generate $\sim\!78$\,\% of the XRB. Using our
slope best estimate of $b = 2.02$, sources producing $2\le k \le
20$ counts contribute further $10$\,\%. Assuming that the
point-like sources generate the whole extragalactic XRB, the $N(S)$
counts should flatten for $S \la 4\cdot 10^{-20}$\,cgs. If some
fraction of the soft XRB is attributed to the diffuse component,
such as the WHIM, the counts flattening has to occur at higher flux
levels. In particular, if the WHIM generates $10$\,\% of the XRB
\citep{soltan07}, the source counts cannot continue with the same
slope below $\sim\!2\cdot 10^{-18}$\,cgs. Apparently, even the
modest extension of the $N(S)$ relationship supplemented with the
precise measurements of the integrated XRB would provide a valuable
data for the investigation of the diffuse component.

In the S band neither the GNL nor our results are consistent with
the predicted counts of AGNs based on a wide class of evolutionary
models (e.g.  \citep{miyaji00}, \citet{gilli01}, \citet{ueda03}).
Consequently, a new population of objects emerging below $\sim
10^{-16}$\,cgs is required.

Young and/or starburst galaxies appear as a natural candidates for
such sources. The XRB spectrum between $1$\,keV and $\sim\!20$\,keV
is adequately approximated by a power law with a photon index of
$1.4$ (\citet{deluca04}, and references therein). Below
$1$\,keV the conspicuous XRB softening is observed
\citep{gilli01}. The soft excess varies from field to field and
evidently exhibits some local and Galactic contribution (e.g.
\citep{markevitch03}). However, the fraction of the soft XRB
generated within the Galaxy is not well established
\citep{soltan07}. Consequently, the exact spectral characteristics
of the extragalactic XRB are not satisfactorily determined. 

A question of the discrete source contribution to the diffuse
background in the radio domain is also present in the literature.
It is interesting that the extragalactic radio source counts exhibit
pronounced slope variations of a character resembling those
observed in the soft X-rays (see \citet{vernstrom11} for the
compilation of the radio data). The counts
derived from the VLA-COSMOS survey at $1.4$\,GHz \citet{bondi08}
above $\sim0.5$\,mJy indicate the slope of $-1.6$,
while just below that flux the slope is equal to $\sim-2.3$.
The counts decline again below $\sim0.1$\,mJy.

In the H band the NNST provides consistent results with the
previous investigations and the source counts do not exhibit any
measurable steepening. It indicates, that the weak sources
generating the count rise in the S band have soft spectra and their
contribution to the XRB above $2$\,keV is low.

\begin{acknowledgements}
I thank all the people generating the Chandra Interactive Analysis
of Observations software for making a really user-friendly
environment.  This work has been partially supported by the Polish
KBN grant 1~P03D~003~27.
\end{acknowledgements}

\end{document}